\documentclass[showpacs,amsmath,amssymb,twocolumn,prl,superscriptaddress]{revtex4-1}

\usepackage{qcircuit}
\usepackage[dvips]{graphicx}
\usepackage{amsmath,amssymb,amsthm,mathrsfs,amsfonts,dsfont}
\usepackage{subfigure, epsfig}
\usepackage{braket}
\usepackage{bm}
\usepackage{enumerate}
\usepackage{enumitem}
\usepackage{color}

\usepackage[colorlinks]{hyperref}
\hypersetup{
	colorlinks=true,
	citecolor = {blue},
    linkcolor=blue,
    filecolor=magenta,      
    urlcolor=cyan,
}

\begin{document}

\title{Error--mitigated digital quantum simulation}
\begin{abstract}
Variational algorithms may enable classically intractable simulations on near-future quantum computers. However, their potential is limited by hardware errors. It is therefore crucial to develop efficient ways to mitigate these errors. Here, we propose a stabiliser-like method which enables the detection of up to 60 -- 80~\% of depolarising errors. Our method is suitable for near-term quantum hardware. Simulations show that our method can significantly benefit calculations subject to both stochastic and correlated noise, especially when combined with existing error mitigation techniques.
\end{abstract}

\date{\today}

\author{Sam McArdle}
\email{sam.mcardle.science@gmail.com}
\affiliation{Department of Materials, University of Oxford, Parks Road, Oxford OX1 3PH, United Kingdom}

\author{Xiao Yuan}
\email{xiao.yuan.ph@gmail.com}
\affiliation{Department of Materials, University of Oxford, Parks Road, Oxford OX1 3PH, United Kingdom}

\author{Simon Benjamin}
\email{simon.benjamin@materials.ox.ac.uk}
\affiliation{Department of Materials, University of Oxford, Parks Road, Oxford OX1 3PH, United Kingdom}

\maketitle

Unlike their classical counterparts, quantum computers can efficiently simulate large quantum systems~\cite{Dirac, Feynman1982, Lloyd1073}. For example, using a quantum computer, we can efficiently find the ground states of systems such as the Fermi-Hubbard model or molecules~\cite{aspuru2005simulated}. Accurately determining the ground states of quantum systems is a first step towards developing new materials~\cite{babbush2017low}, more effective medicines~\cite{Revolution}, and better catalysts~\cite{Reiher201619152}. On first inspection, transformative simulations seem tantalisingly close, requiring only around 100 logical qubits~\cite{Reiher201619152}. However, the required circuit depth necessitates at least $500,000$ physical qubits (based on current error rates and fault-tolerance protocols)~\cite{kivlichan2019trotter,Reiher201619152, babbush2018encoding}, which is many years beyond our current capabilities. 

In contrast, hybrid algorithms, such as the variational quantum eigensolver (VQE)~\cite{peruzzo2014variational, yung2014transistor, VQETheoryNJP}, may not require error correction~\cite{PhysRevA.92.042303}. These algorithms exchange the long gate sequences described above for a polynomial number of short circuits, which dramatically reduces the coherence time required. While previous small experimental demonstrations of the VQE have shown it to be resilient to systematic errors~\cite{PRXH2}, larger experiments have shown that noise can corrupt results~\cite{kandala2017hardware, TrappedIon}. This is perhaps unsurprising; depending on the problem tackled, errors can add or remove particles, effectively changing the system being simulated~\cite{sawaya2016error}.

While error correction is needed to fully suppress these errors, this requires considerable additional resources~\cite{campbell2017roads}. Alternatively, error rates can be lowered by improving the hardware directly. However, this is an enormous challenge; decades of research has reduced error rates to 0.1~\%, but this has not been improved in several years~\cite{ballance2014high, HighFideity16, gaebler2016high}. While efforts to improve error rates continue worldwide, software based methods to effectively reduce error rates are clearly invaluable.

To date, several methods for mitigating errors in near-term quantum hardware have been proposed: the linear~\cite{Li2017, PhysRevLett.119.180509} and exponential~\cite{endo2017practical} extrapolation methods, the quasi-probability method~\cite{endo2017practical, PhysRevLett.119.180509} and the quantum subspace expansion (QSE)~\cite{PhysRevA.95.042308}. Some of these methods have recently been experimentally demonstrated~\cite{kandala2018extending, PhysRevX.8.011021,song2018quasiprobability}. However, these schemes are generally limited to low error rates, where the number of errors expected in the circuit is on the order of unity. There also exist penalty term methods, which drive the calculation towards a state which respects conserved quantities~\cite{VQETheoryNJP,ryabinkin2018symmetry,ryabinkin2019constrained}. While such methods mitigate some algorithmic and coherent errors, they cannot mitigate stochastic errors.

Motivated by these limitations, we have developed a new scheme for error mitigation. Our method uses checks on a suitably constructed trial state to filter errors. It can be used in isolation, or combined with previous error mitigation techniques. We numerically demonstrate our method in an electronic structure calculation on the hydrogen molecule.
The method is low cost, and suitable for the emerging generation of quantum hardware. Our technique is applicable to calculations of both static properties (such as ground and excited states, and vibrational spectra), and dynamical properties (such as time evolved correlation functions). However, herein we focus on the application of our method to the ground state problem, for clarity of exposition.\\


\textit{Conventional VQE.---}  We can use the VQE to find the ground state energy of physical Hamiltonians (see Supplementary Materials).~ 
In molecular simulations, we seek to arrange $N$ electrons among $M$ spin-orbitals, such that the energy of the system is minimised. Here, we use the Jordan-Wigner (JW) mapping, where each qubit represents an electron spin-orbital, the occupation number of which is stored in the $\ket{0}$ or $\ket{1}$ state of the qubit (unoccupied or occupied, respectively). For example, a state describing the hydrogen molecule (H$_2$) is~
\begin{equation}\label{H2wf}
	\ket{\psi^{\mathrm{H_2}}} = \alpha \ket{0101} + \beta \ket{1010} + \gamma \ket{1001} + \delta \ket{0110},
\end{equation}
where $\alpha, \beta, \gamma, \delta$ are complex coefficients. The Hamiltonian can be written as a linear combination of products of Pauli operators, $H = \sum_i g_i \hat{h}_i$, where $g_i$ is a scalar coefficient determining the strength of the term. An example of a typical term is $\hat{h}_i =  X_0Y_1Y_2X_3$.

The VQE, as proposed in Ref.~\cite{peruzzo2014variational}, augments a small quantum processor with a powerful classical computer. The quantum computer is used for classically intractable state preparation, and energy measurement. The state preparation circuit consists of a number of parametrised gates. The circuit used is known as the ansatz, denoted by $U(\vec{\theta})$. This circuit produces a trial state, $\ket{\psi(\vec{\theta})}$. The values of the parameters and the energy of the state they create are input into a classical optimisation algorithm, which seeks the ground state. The energy is calculated by summing the expectation values of each term in the Hamiltonian. To obtain each expectation value, we repeatedly perform the circuit, measure the state produced, and reinitialise. \\



\textit{Stabiliser-VQE.---} In variational simulations, it is often beneficial to initialise the register in a mean-field state, and use a particle-number and spin conserving ansatz~\cite{Barren, particlehole}. This produces states with the correct number of: electrons, $N$, spin-up electrons, $N_\uparrow$, and spin-down electrons, $N_\downarrow$. We refer to these states as `physical' states. As these quantities are conserved, their relevant parity operators are also conserved; $\hat{P}_N \ket{\psi} = P_N \ket{\psi} = (-1)^N \ket{\psi}$ and $\hat{P}_{N_{\uparrow/\downarrow}} \ket{\psi} = P_{N_{\uparrow/\downarrow}} \ket{\psi} = (-1)^{N_{\uparrow/\downarrow}}\ket{\psi}$. This is similar to the concept of stabiliser states used in quantum error correcting codes~\cite{gottesman1997stabilizer}.

There are some ansatze, such as those suggested in Ref.~\cite{particlehole} which are constructed from individual gates which conserve particle number. If a single bit-flip error occurs, it will create or destroy an electron, radically changing the state. For other number and spin conserving ansatze, like the singlet unitary coupled cluster (UCC) ansatz (which, in its canonical form~\cite{romero2017strategies}, is constructed from individual gates which do not necessarily conserve particle number), a single error can propagate and degrade the final state even further. A key concern for the VQE is preserving particle number, as states with electron number far from the true value appear to have a larger energy variance than those with smaller particle number errors~\cite{sawaya2016error}. We present below a method of detecting and removing some of these damaging errors, while still retaining the low qubit resources and gate count of the VQE. \\

In order to detect errors, we introduce an ancilla qubit, and use it to perform measurements of the conserved quantities. When deriving error detection rates, we make the following assumptions: 
\begin{enumerate}[noitemsep]
  \item Errors are symmetric and depolarising.
  \item The error rate is low, such that only one gate malfunctions.
  \item The ansatz circuit is built from individual gates which conserve particle number and spin.
  \item Single qubit gate error rates are negligible compared to two qubit gate error rates.
\end{enumerate}
While our method is still applicable under higher noise rates, different noise models, and using other number conserving ansatze -- as shown by our numerical simulations -- calculating an analytic bound becomes more difficult without these assumptions.\\

\begin{figure}[hbt]
	\begin{align*}
\Qcircuit @C=0.6em @R=.4em {
\lstick{\uparrow : \ket{x_0}}&\qw& \multigate{1}{U(\vec{\theta})}&\qw &\ctrl{2}&\qw&\qw&\qw&\qw&\qw&\meter \\
\lstick{\downarrow : \ket{x_1}}&\qw& \ghost{U(\vec{\theta})}&\qw &\qw&\ctrl{1}&\qw&\qw&\qw&\qw&\meter \\
\lstick{\ket{0}_a}&\qw& \qw&\qw&\targ&\targ&\qw&\qw&\qw & \meter\\
}
\end{align*}
\caption{A circuit to check the particle number parity of a physical trial state. The ancilla should be measured in $\ket{\frac{1}{2} (1- (-1)^{N})}$. If errors occur, and the measured value of the ancilla is not correct, the measurement of the Hamiltonian term $\hat{h}_i$ on the register is not performed, and the circuit is reinitialised.
}\label{parityCheck}
\end{figure}
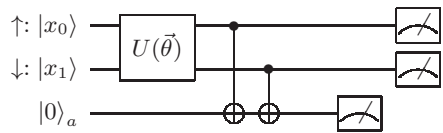

The most simple check is of the total electron number parity. This procedure is shown in Fig.~\ref{parityCheck}. Each CNOT gate flips the ancilla qubit if the corresponding spin-orbital is occupied by an electron. The circuit enables the detection of any error which changes the electron number parity by one. Under the assumptions described above, we are able to detect and remove 53~\% of error events (8/15 errors in the depolarising noise model, as described in the Supplementary Materials).~
In the Supplementary Materials we present an alternative circuit for this parity check, which can reduce the impact of qubit readout errors. The qubit readout error rate is currently around 1~\% for superconducting qubits~\cite{preskill2018quantum}. This alternative circuit also makes it possible to combine our method with existing variational algorithms for real~\cite{Li2017} and imaginary~\cite{mcardle2018variational} time evolution.

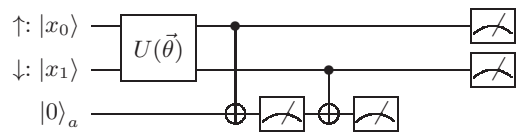
\begin{figure}[hbt]
	\begin{align*}
\Qcircuit @C=0.6em @R=.4em {
\lstick{\uparrow : \ket{x_0}}&\qw& \multigate{1}{U(\vec{\theta})}&\qw &\ctrl{2}&\qw&\qw&\qw&\qw&\qw&\qw&\qw&\meter \\
\lstick{\downarrow : \ket{x_1}}&\qw& \ghost{U(\vec{\theta})}&\qw&\qw&\qw &\ctrl{1}&\qw&\qw&\qw&\qw&\qw&\meter \\
\lstick{\ket{0}_a}&\qw& \qw&\qw&\targ&\meter&\targ&\meter \\
}
	\end{align*}
\caption{A circuit to measure the spin parities. We compute the spin-up parity onto the ancilla, and measure it. We then reset the ancilla to $\ket{0}$, and measure the spin-down parity.}\label{parity2}
\end{figure}

To increase the proportion of errors detected, we can perform the circuit shown in Fig.~\ref{parity2} which measures both the spin-up parity and spin-down parity. This enables us to detect additional two qubit bit flip errors that we were unable to detect using the single parity check above. We can effectively filter out 66~\% of errors, as shown in the Supplementary Materials).

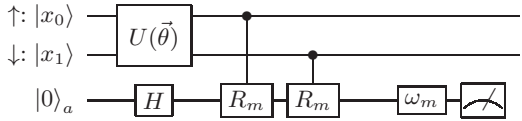
\begin{figure}[hbt]
	\begin{align*}
\Qcircuit @C=0.6em @R=.7em {
\lstick{\uparrow : \ket{x_0}}&\qw& \multigate{1}{U(\vec{\theta})}&\qw &\ctrl{2}&\qw&\qw&\qw&\qw&\qw&\qw&\qw \\
\lstick{\downarrow : \ket{x_1}}&\qw& \ghost{U(\vec{\theta})}&\qw &\qw&\ctrl{1}&\qw&\qw&\qw&\qw&\qw&\qw \\
\lstick{\ket{0}_a}&\qw& \gate{H}&\qw&\gate{R_m}&\gate{R_m}&\qw&\qw&\qw & \gate{\omega_m} & \meter \\
}
\end{align*}
\caption{The circuit which measures the $m$\textsuperscript{th} bit of the electron number, $N_m$. The $R_m$ gates are given by $\mathrm{diag}(1, e^{\pi i / 2^{m-1}})$. The gate $\omega_m$ is given by $\mathrm{diag}(1, e^{- \mathrm{dec}(N_{m-1} \dots N_1) \pi i / 2^{m-1}})$, where $\mathrm{dec}(N_{m-1} \dots N_1)$ is the decimal representation of the binary string $N_{m-1} \dots N_1$. We define $\omega_1$ as the identity matrix. Measurement of the ancilla is in the $X$ basis.}\label{conserved}
\end{figure}

We can also measure the electron number and spin numbers directly, using an iterative procedure. We first write the electron number in binary. We then use the circuit in Fig.~\ref{conserved} to measure the first bit in $N$, by setting $m=1$. We denote the bit value measured as $N_1$. We then repeat the circuit in Fig.~\ref{conserved} to measure the second bit in $N$, by setting $m=2$, and using our measurement of $N_1$ in the rotation $\omega_2$. In general, we can measure the $m$\textsuperscript{th} bit of $N$ using the circuit in Fig.~\ref{conserved}, constructing $\omega_m$ using our measurements of the $m-1$ preceeding bits in $N$. When no errors have occured, the ancilla is in the state 
\begin{equation}\label{numbermeasure}
\ket{\phi} = \frac{1}{\sqrt{2}} (\ket{0_a} + e^{N_m \pi i} \ket{1_a} ).
\end{equation}
If $N_m = 0$ we measure the ancilla in $\ket{+}$, while if $N_m = 1$ we measure the ancilla in $\ket{-}$. A worked example for $N=3$ is given in the Supplementary Materials.~
We can measure the spin numbers using a similar procedure.

The exact fraction of errors that can be detected will depend on the trial state produced by the ansatz, but has a maximum value of 80~\% (see Supplementary Materials).~
In total, $M \mathrm{log}_2(N)$ control gates are needed to measure the electron number, or both spin numbers.\\

While our method can filter a fraction of the possible errors, the results will not be completely noise free. Consequently, we can combine our technique with other methods of error mitigation to further improve accuracy. The most straightforward approach is to use the detection method to filter out errors, and then use the extrapolation technique to obtain highly accurate results.\\

\textit{Hardware implementation.---} We now consider how to implement our technique in current leading architectures; specifically trapped ion and superconducting systems. The circuit structure required is similar to that of a stabiliser evaluation for a topological code, which has been investigated for both platforms~\cite{bermudez2017assessing, fowler2012surface}. As can be seen from Fig.~\ref{parityCheck}, the optimal implementation requires non-local gates. Trapped ion systems can perform gates between non-adjacent ions~\cite{linke2017experimental}, so these circuits do not present any particular difficulty. Moreover, modular architectures are feasible designs for trapped ion systems~\cite{monroe2014large,nickerson2014scalable}.
Such architectures are network-like and could be constructed with the connectivity required for our circuits. As the coherence times of trapped ion qubits are considerably longer than their readout times~\cite{noek2013highspeed}, it is possible to carry out the checks using a single ancilla that is repeatedly measured and reinitialised.

In contrast, superconducting qubits typically have more limited connectivity, that may be nearest-neighbour. With such an architecture, we can implement our particle number check by using $O(M)$ SWAP gates to move the ancilla along the qubit register. It is possible to realise the parity checks with a shorter circuit, which was discussed in Ref.~\cite{bonet2018mitigation}, and which we reproduce in the Supplementary Materials. In contrast, the number of gates required for a general UCC ansatz is $O(M^3)$~\cite{motta2018low}. For physical systems of interest, requiring $M=50-100$ qubits, the gate count will be dominated by the ansatz circuit. Consequently, we expect the additional gates to have little impact on the detection rates derived above. For superconducting qubits, the measurement time can be of a comparable order of magnitude to the coherence time~\cite{wendin2017quantum}. As such, it may be preferable to use multiple ancilla qubits, rather than to repeatedly reinitialise a single ancilla. This modest overhead constitutes two ancilla qubits for the spin number parity check, and $\mathrm{log_2}N$ ancilla qubits for the spin number check. \\

\textit{Results.---} We tested our method's efficacy in a VQE calculation on the simplest model of H$_2$, with two electrons in four spin-orbitals. We used a spin-conserving UCC ansatz applied to the Hartree-Fock state. We did not consider the parameter update step of the VQE, so as to examine the effect of errors without consideration of a classical optimisation algorithm~\cite{sawaya2016error}. Numerical simulations were performed using QuEST~\cite{jones2018quest}, and simulation code can be found at Ref.~\cite{myGithub}. Our aim was to measure the energy to within `chemical accuracy' ($1.6$~mHartree), which enables the prediction of reaction rates to within an order of magnitude at room temperature.

To detect errors, we performed error-prone checks of both the spin-up and spin-down parity numbers, using the circuit shown in the Supplementary Materials. This circuit has nearest-neighbour connectivity, and so lower bounds the efficacy of our method. We designed our simulations to mimic the actions of an experimentalist; the expectation value of each term in the Hamiltonian was found by repeating the circuit and measurement procedure many times. The number of measurements used is discussed in the Supplementary Materials. \\

\begin{figure}[h]\centering
\includegraphics[width=1.0\columnwidth]{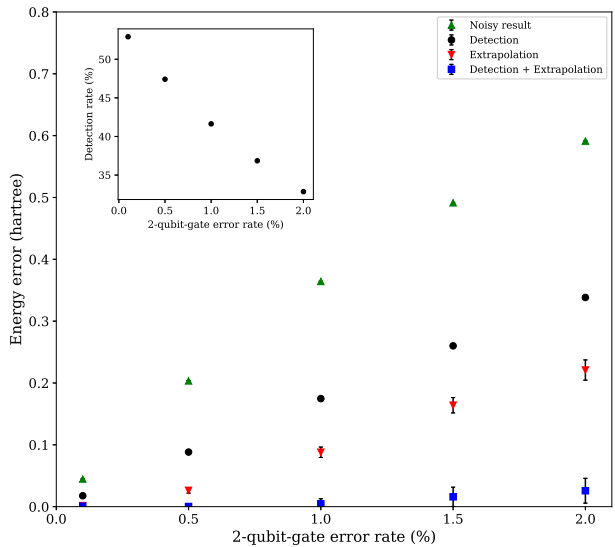}
 \caption{Comparing methods of mitigating errors in simulations of H$_2$. The detection rates shown in the inset were obtained from the numerical simulations. The true energy value was $-1.1227$~Hartree. The error bars upper bound the standard error in the result.} \label{ErrorComp}
\end{figure}

Initially, we considered energy measurements on a trial state that contained all four possible vectors, as described by Eq.~(\ref{H2wf}). We used a symmetric depolarising noise model, and set the two qubit gate error rate to be 10 times larger than the single qubit gate error rate. We measured the energy of the state prepared by the ansatz under the following conditions:
\begin{enumerate}[noitemsep]
  \item No parity check, Errors, No extrapolation.
  \item No parity check, Errors, Extrapolation.
  \item Parity check, Errors, No extrapolation.
  \item Parity check, Errors, Extrapolation.
\end{enumerate}
We compare these results to the true energy extracted in the limit of infinite measurements and no gate errors. The results are shown in Fig.~\ref{ErrorComp}, using two qubit gate error rates ranging from 0.1~\% to 2~\%. There were 92 single qubit gates and 56 two qubit gates in the UCC ansatz circuit -- which we can approximate as 65 two qubit gates. The parity checks contributed an additional 8 error-prone two qubit gates.

We see from Fig.~\ref{ErrorComp} that the error detection method alone improves the accuracy of our energy measurements. While detection is less effective than extrapolation, a greater benefit can be obtained by combining the two methods. The accuracy of the combined method does not worsen significantly as the error rate increases, unlike the two individual methods. We observe from the inset plot that the fraction of detected errors falls approximately linearly with increasing error rate. When the error rate is small we detect around 53~\% of errors. If we assume that we can only detect errors which occur during the ansatz circuit (due to the use of nearest-neighbour gates for the parity checks) then our detection probability is given by the probability of an error happening in the ansatz circuit ($65/73$), multiplied by the probability of detection ($10/15$), which is roughly $59~\%$. We attribute the deviation of our result from this value to the use of a UCC ansatz, which enables errors to propagate. At higher error rates, multiple errors are able to occur in the circuit, which reduces the fraction of errors that we can detect to close to one third. \\

We also used our method when calculating the dissociation curve of H$_2$. We compare the true energy values with: noisy measurements without error mitigation, noisy measurements with extrapolation, and noisy measurements with detection and extrapolation. The two qubit depolarising error rate was $0.1~\%$, which has been achieved in a controlled setting~\cite{HighFideity16, gaebler2016high}, and should be targeted in near-future quantum computers. We combined this with temporally correlated over/under rotations of up to 1~\%, (see Supplementary Materials). While previous experimental VQE calculations on H$_2$ have achieved accurate results with higher error rates~\cite{PRXH2, kandala2017hardware, PhysRevX.8.011021, kandala2018extending, TrappedIon, ganzhorn2018gateefficient}, these experiments used fewer gates (and often fewer qubits) -- obtained using simplifications that are applicable to H$_2$, but not to larger molecules. We chose to use the non-optimised circuits to ensure that our simulations were representative of general chemistry problems.

\begin{figure}[h]\centering
\includegraphics[width=1.0\columnwidth]{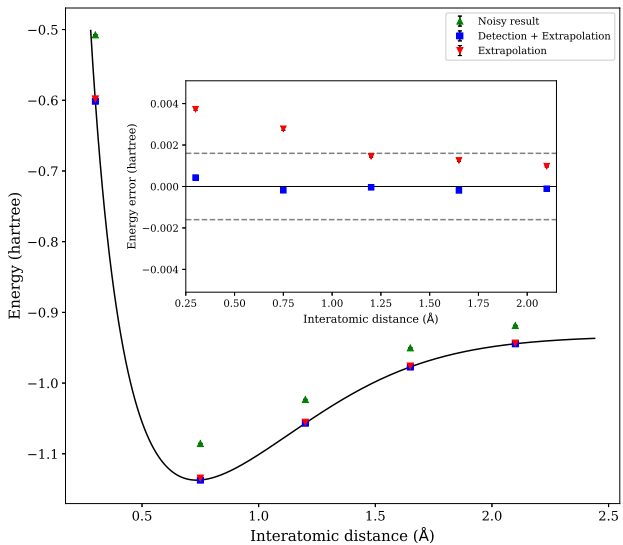}
 \caption{Comparing methods of mitigating errors in simulations of H$_2$. The error model is described in the main text. The inset shows the residual from the true value. The dashed lines in the inset mark chemical accuracy ($\pm$1.6~mHartree). The error bars show the true standard error in the result, given in the Supplementary Materials.} \label{ThirdTest}
\end{figure}

We see from Fig.~\ref{ThirdTest} that our method can obtain chemically accurate energies, even when the results would otherwise be corrupted by noise. The combined mitigation method achieves quantitatively accurate results, with a mean absolute residual of $0.2$~mHartree. Compared to the unmitigated results, the deviation from the true value is reduced by a median factor of 239 (with a range of 141 to 818, and a mean of 340) by combining the error mitigation techniques. Compared to the extrapolated results, the deviation from the true value is reduced by a median factor of 9.1 (with a range of 6.6 to 35.6, and a mean of 15.0) by combining the two mitigation methods. As energies are exponentiated when calculating reaction rates, this improvement will be magnified when performing calculations of interest.\\

\textit{Discussion.---} We have introduced a method to mitigate errors in near-term digital quantum simulations, which requires minimal additional resources. Our technique can be used to detect errors which change conserved quantities. It can be applied to calculations of both static and dynamical properties.

Our method can improve the accuracy of variational calculations, especially when combined with the extrapolation method of error mitigation. We simulated a noisy VQE calculation of the hydrogen molecule, and found that using this approach reduced the deviation from the true result by two orders of magnitude. Our simulations were performed with nearest-neighbour connectivity, showing the method's practicality.

Recent work has shown that surpassing classical simulation techniques with non-error-corrected quantum computers will likely require at least $10^4$ gates~\cite{motta2018low, jiang2018hubbard,mcardle2018review}. However, it is difficult to foresee error rates below 0.01~\% -- at which point, we would expect an error to occur in every circuit, on average. Error mitigation techniques, such as those presented herein, may enable us to extract meaningful results from these simulations, providing a practical use for near-term quantum hardware. Consequently, implementing these techniques on recently announced devices~\cite{wright2019benchmarkingionq, google2018bristlecone, ibm2019one, nersisyan2019manufacturing} -- which will be cloud-accessible, have two-qubit-gate fidelities approaching 99~\%, and possess tens of qubits -- would provide an interesting avenue for future study.

While our method can detect a large proportion of errors, additional mechanisms will be required to provide error resilience for long circuits on non-error corrected machines. One possibility is to combine our method with a two qubit phase-error detection code. Alternatively, one could utilise other invariant quantities. It was noted in Ref.~\cite{bravyi2017tapering} that the Hamiltonians of small molecules contain several symmetries. Evolution under a Hamiltonian variational ansatz may conserve these quantities, enabling the design of additional checks. Future work will investigate the performance of concatenated mitigation methods for larger problems. \\ \\

\textit{Acknowledgements.---} \cite{referencesNote, BravyiKitaev12, nielsen2002quantum, ekert2002direct, kitaev1995phase, openfermion, whitfield2011simulation, hughesandhase} 
We thank S.~Endo for insightful discussions. This work was supported by BP plc and the EPSRC National Quantum Technology Hub in Networked Quantum Information Technology (EP/M013243/1). We acknowledge the use of the University of Oxford Advanced Research Computing facility. \\ 

\textit{Note added.---} After this paper was released, a relevant preprint was posted by Bonet-Monroig~\textit{et al.}~\cite{bonet2018mitigation}. They suggest a similar method of error mitigation, and an elegant method for error mitigation via post-processing. While their results focus on comparing their techniques, our results are consistent and can be compared.

\section{Supplementary Materials}

\tableofcontents

\section{Digital quantum simulation}\label{compchem}
We can use the variational quantum eigensolver (VQE) to find the ground state energy of a physical Hamiltonian~\cite{peruzzo2014variational}. Following most of the previous work on near-term quantum simulation, we use the second quantised formalism. In this section, we focus primarily on the simulation of molecules, although the procedure is similar for lattice based models such as the Fermi-Hubbard model. We project the Hamiltonian onto a finite number of basis wavefunctions, $\{{\phi_p}\}$, which approximate spin-orbitals. Electrons are excited into, or de-excited out of, these orbitals by fermionic creation ($a_p^\dag$) or annihilation ($a_p$) operators, respectively. These operators obey fermionic anti-commutation relations, which enforce the antisymmetry of the wavefunction, a consequence of the Pauli exclusion principle. In the second quantised representation, the electronic Hamiltonian is written as
\begin{equation}\label{Eq:FH}
	H = \sum_{p,q}h_{pq}a^\dag_p a_q + \frac{1}{2}\sum_{p,q,r,s}h_{pqrs}a^\dag_p a^\dag_q a_ra_s,
\end{equation}
with 
\begin{equation}
	\begin{aligned}\label{Integrals}
		h_{pq}&=\int d\textbf{x} \phi_p^*(\textbf{x}) \left(-\frac{\nabla^2_i}{2} -\sum_{I}\frac{Z_I}{|\mathbf{r}-\mathbf{R}_I|}\right) \phi_q(\mathbf{x}),\\
		 h_{pqrs}&=\int d\mathbf{x}_1 d\mathbf{x}_2\frac{\phi_p^*(\mathbf{x}_1) \phi_q^*(\mathbf{x}_2)\phi_s(\mathbf{x}_1)\phi_r(\mathbf{x}_2)}{|\mathbf{r}_1-\mathbf{r}_2|},
	\end{aligned}
\end{equation}
where $\mathbf{x}$ is a spatial and spin coordinate, and $\mathbf{R}_I$ is the position of the $I$\textsuperscript{th} nucleus. \\

This fermionic Hamiltonian can be mapped to a qubit Hamiltonian by employing an encoding method. The two most common methods are the Jordan-Wigner (JW) and Bravyi-Kitaev (BK) mappings~\cite{BravyiKitaev12}. In this work, we use the JW encoding, as it was found in Ref.~\cite{sawaya2016error} that the BK mapping was more susceptible to errors than JW mapped states. In the JW encoding, we store the occupation number of an orbital in the $\ket{0}$ or $\ket{1}$ state of a qubit (unoccupied or occupied, respectively). The mapping between the fermionic creation and annihilation operators and qubit gates is given by~\cite{BravyiKitaev12}
\begin{equation}
	\begin{aligned}
		a_p &= (X_p + iY_p)\otimes Z_{p-1}\otimes \dots\otimes  Z_{0},\\
		a_p^\dag &= (X_p - iY_p) \otimes Z_{p-1}\otimes \dots\otimes  Z_{0}.
	\end{aligned}
\end{equation}
The $X$ and $Y$ operators change the occupation number of the target orbital, while the string of $Z$ operators enforces electron exchange antisymmetry. The JW mapped Hamiltonian of a molecule can be written as a linear combination of products of Pauli operators,
\begin{equation}
	H = \sum_j^M g_j \hat{h}_j = \sum_j^M g_j \prod_i \sigma_i^j,
\end{equation}
where $g_j$ are coefficients, $\sigma_i^j$ represents one of $I$, $X$, $Y$,  or $Z$, $i$ denotes which qubit the operator acts on, and $j$ denotes which term in the Hamiltonian we apply. For example,
\begin{equation}\label{VQE_Hamil_spec}
	\hat{h}_j =  X_0Y_1Y_2X_3Z_4Z_5.
\end{equation}
As each term in the fermionic Hamiltonian contains an even number of creation and annihilation operators, each term in the qubit Hamiltonian will contain an even number of $X$ or $Y$ operators, and thus will conserve particle number parity.

\section{Noise models}

\subsection{Depolarising noise}
We consider a symmetric depolarising channel in all quantum gates. The single qubit gate channel is given by~\cite{nielsen2002quantum}
\begin{equation}
    \mathcal{E}(\rho) = (1-p)\rho + \frac{p}{3}(X\rho X + Y\rho Y + Z \rho Z),
\end{equation}
where $p$ is the probability that the gate malfunctions. The two qubit gate depolarising channel is given by
\begin{equation}
    \mathcal{E}(\rho) = (1-p)\rho + \frac{p}{15}\sum_{i,j} (O_{i,1} O_{j,2} \rho O_{i,1}^\dag O_{j,2}^\dag ),
\end{equation}
where $O_{i, q}$ is the operator $O_{i}$ acting on qubit $q$, and $O_i$ runs over $I, X, Y, Z$, and the possibility $O_i = O_j = I$ is not included (this possibility, of no error occurring, is described by the term outside of the sum).

\subsection{Temporally correlated over/under rotations}

For our second error model, we have considered temporally correlated over/under rotations. Each discrete gate in the circuit (eg. Hadamard, CNOT, etc) is replaced by a parametrised equivalent, eg. the Hadamard gate $H$ was replaced with
\begin{equation}
    H = \\
    \begin{bmatrix}
    \mathrm{cos}(\epsilon \frac{\pi}{4}) & \mathrm{sin}(\epsilon \frac{\pi}{4}) \\
    \mathrm{sin}(\epsilon \frac{\pi}{4})  & -\mathrm{cos}(\epsilon \frac{\pi}{4}) 
    \end{bmatrix}
\end{equation}
Similarly, the parametrised CNOT gate is represented by
\begin{equation}
    \mathrm{CNOT} = \\
    \begin{bmatrix}
    1 & 0 & 0 & 0 \\
    0 & 1 & 0 & 0 \\
    0 & 0 & 
    \mathrm{icos}(\epsilon \frac{\pi}{2}) & \mathrm{sin}(\epsilon \frac{\pi}{2}) \\
    0 & 0 & 
    \mathrm{sin}(\epsilon \frac{\pi}{2})  & \mathrm{icos}(\epsilon \frac{\pi}{2}) 
    \end{bmatrix}
\end{equation}
All rotation gates $R(\theta)$ were replaced with $R(\epsilon \theta)$, except for those used to change the measurement basis when measuring a Hamiltonian term. If the parameter $\epsilon$ is set to 1, then the original gates are recovered. On each iteration of the circuit, we generated a
random $\delta$ for each qubit, $\delta_q$, uniformly distributed between $\pm 0.01$.  For all of the single qubit gates acting on qubit $q$ in the circuit, we set $\epsilon_q = 1 - \delta_q/10$. For the target qubit of two qubit gates, we set $\epsilon_q = 1 - \delta_q$. As such, $0.99 < \epsilon_q < 1.01$. In each instance of the circuit each qubit receives the same over or under rotation for each of the gates where it is the target (we chose that control qubits do not suffer an over/under rotation in CNOT gates). This represents a strongly temporally correlated noise model. \\

\section{Error detection rates} \label{detectrates}
When deriving error detection rates, we make the following assumptions: 
\begin{enumerate}[noitemsep]
  \item The errors are symmetric and depolarising.
  \item The error rate is low, such that only one gate malfunctions in the circuit.
  \item The ansatz circuit is built from individual gates which conserve particle number and spin.
  \item Single qubit gate error rates are negligible compared to two qubit gate error rates.
\end{enumerate}
While our method is still applicable under higher noise rates, different noise models, and using other number conserving ansatze (such as the unitary coupled cluster ansatz used in our numerical simulations), an analytic bound on the error detection rate becomes more difficult without these assumptions.\\

\subsection{Total electron number parity check}

By performing a single parity check of the total electron number, we are able to detect 53~\% of errors under the above assumptions. In the symmetric depolarising noise model the following errors are equally likely following a two qubit gate between qubits $i$ and $j$:
\begin{gather}
X_iI_j, \quad I_iX_j, \quad Y_iI_j, \quad I_iY_j, \quad X_iZ_j, \quad Z_iX_j, \quad Y_iZ_j, \quad Z_iY_j,  \\
Z_iI_j, \quad I_iZ_j, \quad X_i X_j, \quad Y_i Y_j, \quad Z_i Z_j, \quad X_iY_j, \quad Y_iX_j. \notag
\end{gather}

We can see that all of the errors in the top row change the electron number parity of a state vector. As a result, using a single parity check, we can detect 8/15 $\approx$ 53~\% of these errors.

This is true both if the error occurs during the ansatz circuit (we can detect $X_iI_j, I_iX_j, Y_iI_j, I_iY_j, X_iZ_j, Z_iX_j, Y_iZ_j, Z_iY_j$ out of the 15 possible errors on two of the register qubits) or during the parity check gate sequence (assuming that we use non-local gates for the parity check, then we can detect $X_rX_a, X_rY_a, Y_rX_a, Y_rY_a, Z_rX_a, Z_rY_a, I_rX_a, I_rY_a$, where $r$ and $a$ denote register and ancilla qubits, respectively).

\subsection{Spin-up and spin-down parity check}
Using the spin-up and spin-down parity checks, we can detect additional errors beyond the 8/15 detectable using the electron number parity measurement. We will be able to detect certain two qubit errors, which change the value of either spin-parity. However, we will not be able to detect all two qubit bit-flip errors. 

There are ${M} \choose {2}$~$= \frac{M(M-1)}{2}$ ways of distributing a two qubit bit-flip error between the $M$ spin-orbitals. We are unable to detect errors which both occur on the spin-up orbitals, or both on spin-down orbitals. We are able to detect errors which occur on one orbital of each spin. There are ${M/2} \choose {1}$~$= M/2$ ways of distributing an error amongst half of the orbitals. As a result, there are $\frac{M^2}{4}$ errors we can detect.

For a 2 spin-orbital system, we can detect $\frac{2^2/4}{(2 \times 1)/2} = 100$~\% of double bit-flip errors.
For a 4 spin-orbital system, this reduces to 66~\% of double bit-flip errors.

When $M$ is large, we can detect $\frac{M^2/4}{M(M-1)/2} \approx \frac{M^2/4}{M^2/2} = \frac{1}{2}$ of double bit-flip errors. As there are four possible types of double bit-flip errors, and we can detect half of the occurences of each of them, this effectively increases the number of error events we can detect by 2/15, to 10/15 $\approx$ 66~\%.\\

The above calculation holds when the error occurs during the ansatz circuit. If the error happens during the error detection procedure, the situation is more complicated. We can detect around 66~\% of errors by performing the total electron number parity check first, then the spin-up and then the spin-down parity checks. First assume that the error happens during the first parity check. Using the total parity check, we can detect 8/15 errors that occur (we can detect $X_rX_a, X_rY_a, Y_rX_a, Y_rY_a, Z_rX_a, Z_rY_a, I_rX_a, I_rY_a$, where $r$ and $a$ denote register and ancilla qubits, respectively). Using the spin-up parity check, we are able to detect any of $X_r I_a, X_r Z_a, Y_r I_a, Y_r Z_a$ that happen to one of the spin-up register qubits. The same is true for errors on the spin-down register qubits. As a result, we can detect all $X_r I_a, X_r Z_a, Y_r I_a, Y_r Z_a$ errors resulting from the first parity check, in addition to those described above. This is 12/15 of the errors in the depolarising noise model. If the error occurs during either of the other two parity checks, we are only able to detect the 8/15 errors listed above. However, as the total electron number parity check requires twice as many gates as either of the spin-up or spin-down parity checks, the error is equally like to occur in the first parity check as in either of the other checks. As such, on average we can detect $10/15 \approx 66~\%$ of errors, even if they occur in the error checking process. Alternatively, given that the ansatz circuit will contain many more two qubit gates than the parity check circuits, we can just use the spin-up and spin-down parity checks, and assume that it is far more likely that the error will occur in the ansatz circuit.

\subsection{Electron number check}
\begin{figure}[hbt]
	\begin{align*}
\Qcircuit @C=0.6em @R=.7em {
\lstick{\ket{x_0}}&\qw& \multigate{5}{U(\vec{\theta})}&\qw &\ctrl{6}&\qw&\qw&\qw&\qw&\qw&\qw&\qw&\qw&\qw \\
\lstick{\ket{x_1}}&\qw& \ghost{U(\vec{\theta})}&\qw &\qw&\ctrl{5}&\qw&\qw&\qw&\qw&\qw&\qw&\qw&\qw \\
\lstick{\ket{x_2}}&\qw& \ghost{U(\vec{\theta})}&\qw& \qw&\qw&\ctrl{4}&\qw&\qw&\qw&\qw&\qw&\qw&\qw \\
\lstick{\ket{x_3}}&\qw& \ghost{U(\vec{\theta})}&\qw& \qw&\qw&\qw&\ctrl{3}&\qw&\qw&\qw&\qw&\qw&\qw \\
\lstick{\ket{x_4}}&\qw& \ghost{U(\vec{\theta})}&\qw&\qw&\qw& \qw&\qw&\ctrl{2}&\qw&\qw&\qw&\qw&\qw \\
\lstick{\ket{x_5}}&\qw& \ghost{U(\vec{\theta})}&\qw&\qw&\qw& \qw&\qw&\qw&\ctrl{1}&\qw&\qw&\qw&\qw \\
\lstick{\ket{0}_a}&\qw& \gate{H}&\qw&\gate{R_1}&\gate{R_1}&\gate{R_1}&\gate{R_1}&\gate{R_1}&\gate{R_1}&\qw & \meter \\
}
\end{align*}
\caption{The circuit which measures the $1$\textsuperscript{st} bit of the electron number, $N_1$. The $R_1$ gates are given by $\mathrm{diag}(1, e^{\pi i})$. Measurement of the ancilla is in the $X$ basis.}\label{numberexample1}
\end{figure}
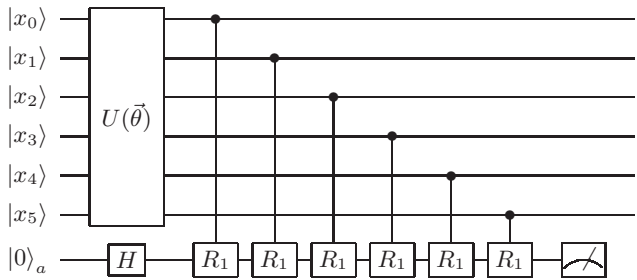

\begin{figure}[hbt]
	\begin{align*}
\Qcircuit @C=0.6em @R=.7em {
\lstick{\ket{x_0}}&\qw& \multigate{5}{U(\vec{\theta})}&\qw &\ctrl{6}&\qw&\qw&\qw&\qw&\qw&\qw&\qw&\qw&\qw \\
\lstick{\ket{x_1}}&\qw& \ghost{U(\vec{\theta})}&\qw &\qw&\ctrl{5}&\qw&\qw&\qw&\qw&\qw&\qw&\qw&\qw \\
\lstick{\ket{x_2}}&\qw& \ghost{U(\vec{\theta})}&\qw& \qw&\qw&\ctrl{4}&\qw&\qw&\qw&\qw&\qw&\qw&\qw \\
\lstick{\ket{x_3}}&\qw& \ghost{U(\vec{\theta})}&\qw& \qw&\qw&\qw&\ctrl{3}&\qw&\qw&\qw&\qw&\qw&\qw \\
\lstick{\ket{x_4}}&\qw& \ghost{U(\vec{\theta})}&\qw&\qw&\qw& \qw&\qw&\ctrl{2}&\qw&\qw&\qw&\qw&\qw \\
\lstick{\ket{x_5}}&\qw& \ghost{U(\vec{\theta})}&\qw&\qw&\qw& \qw&\qw&\qw&\ctrl{1}&\qw&\qw&\qw&\qw \\
\lstick{\ket{0}_a}&\qw& \gate{H}&\qw&\gate{R_2}&\gate{R_2}&\gate{R_2}&\gate{R_2}&\gate{R_2}&\gate{R_2}&\gate{\omega_2}&\qw & \meter \\
}
\end{align*}
\caption{The circuit which measures the $2$\textsuperscript{nd} bit of the electron number, $N_2$. The $R_2$ gates are given by $\mathrm{diag}(1, e^{\pi i/2})$. The $\omega_2$ gate is given by $\mathrm{diag}(1, e^{-\mathrm{dec}(N_1) \pi i/2})$, where $\mathrm{dec(N_1)}$ is the decimal value of the first bit of $N$. Measurement of the ancilla is in the $X$ basis.}\label{numberexample2}
\end{figure}
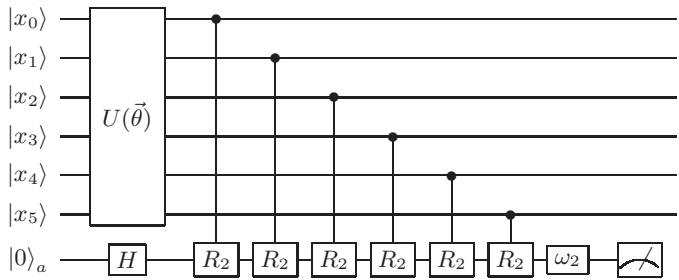

We provide a worked example for the case of a system with $N=3$ electrons in $M=6$ orbitals. We first apply the circuit shown in Fig.~\ref{numberexample1}
After this circuit, if no errors have occured, the ancilla is in the state
\begin{equation}
\ket{\phi} = \frac{1}{\sqrt{2}} (\ket{0_a} + e^{3 \pi i} \ket{1_a} ) = \ket{-}.
\end{equation}
Measuring the ancilla in the $X$ basis informs us that the first bit in $N$ is 1. We then apply the circuit shown in Fig.~\ref{numberexample2}. The $\omega_2$ gate is given by $\mathrm{diag}(1, e^{-\mathrm{dec}(N_1) \pi i/2}) = \mathrm{diag}(1, e^{- \pi i/2})$.

The state of the ancilla is
\begin{equation}
\ket{\phi} = \frac{1}{\sqrt{2}} (\ket{0_a} + e^{\pi i} \ket{1_a} ) = \ket{-}.
\end{equation}
This informs us that the second bit in $N$ is also 1. As a result, we know the number of electrons in the state is 3. If a different value were measured, this would inform us that an error has occured, and therefore that the result should be discarded. \\

It is only possible to naively estimate the maximum error detection rate using the particle number checks, for reasons which are discussed in more detail below. We can initially perform the parity measurements described above to detect 66~\% of errors, under the assumptions listed at the beginning of this section. We then make the assumption that the wavefunction consists of only one Slater determinant. We also assume that the error occurs during the ansatz circuit, and not during the number measurement procedure. 

We are unable to detect any two qubit bit-flip error which acts on one occupied orbital and one unoccupied orbital of the same spin, as this mimics a spin conserving excitation operator. If we consider the spin-up electron number, then the number of possible two qubit bit flip errors which act on the occupied spin-up orbitals is ${N/2} \choose {2}$ (assuming there are an equal number of spin-up and spin-down electrons, without loss of generality). The number of possible two qubit bit flip errors which act on the unoccupied spin-up orbitals is ${M/2-N/2} \choose {2}$. The total number of possible two qubit errors (where both errors happen on the spin-up orbitals) is ${M/2} \choose {2}$. As such, the detection rate is

\begin{equation}
\frac{\frac{N}{2}(\frac{N}{2} -1) + \frac{M-N}{2}(\frac{M-N}{2} -1)}{\frac{M}{2}(\frac{M}{2} -1)}.
\end{equation}

In the limit that $M \rightarrow \infty$ and $M >> N$, all errors occur on the unoccupied orbitals, and so change the electron number and spin numbers. As such, we are able to detect all two qubit bit flip errors. This means we are able to detect up to 12/15 = 80~\% of single gate errors in the depolarising noise model (although this is clearly unattainable in practice). \\

However, this only applies to a single Slater determinant, while the output of our ansatz will likely be a superposition of multiple determinants. The reason for this approximation is that double bit-flip errors may change the electron and spin numbers of certain determinants, while leaving others unchanged. For example, consider the state

\begin{equation}
\ket{\psi} = \frac{1}{\sqrt{2}}(\ket{001011} + \ket{100101},
\end{equation}

which has an electron number of 3, a spin up number of 2, and a spin down number of 1 (the rightmost 3 orbitals are spin-up, the leftmost 3 orbitals are spin-down). If this state undergoes a double bit flip error given by $X_0X_1$, the state becomes

\begin{equation}
\ket{\psi'} = \frac{1}{\sqrt{2}}(\ket{001000} + \ket{100110}.
\end{equation}

This new state has the correct spin up and spin down parities, so the error is undetectable using parity checks. Moreover, we see that while the first Slater determinant in the wavefunction has an incorrect particle number, the second still has the correct particle number. If we apply the electron number measuring scheme described above, we will still measure 1 for the first bit in the electron number. However, when measuring the second bit, there is a 50~\% chance we measure 1 (ancilla in $\ket{-}$), and a 50~\% chance we measure 0 (ancilla in $\ket{+}$). If we measure 0, we conclude that an error has occured, and correctly discard the state. However, if we measure 1, we conclude that the electron number is correct, and no error has occured. As a result, we collapse the wavefunction into the state
\begin{equation}
\ket{\psi''} = \ket{100110},
\end{equation}
and measure the energy of this state, which may be far in energy from the state which the ansatz intended to produce. As such, the true rate of accepting uncorrupted states is below the 80~\% maximum discussed above, but impractical to calculate analytically.

\subsection{Non-violation of the variational principle}
In this section, we show that the measured energy still obeys the variational principle after the filtering process described above, such that energy values below the true ground state energy cannot be measured. We can prove this by considering the Hamiltonian produced after mapping to qubits. The physical states that our ansatz generates correspond to a small region of the full Hilbert space. Errors may take the state produced outside of this region. Under a general noise channel, the density matrix for the system is given by
\begin{equation}
    \rho = \sum_i p_i \ket{\psi_i}\bra{\psi_i},
\end{equation}
where $p_i$ is the probability of the register being in state $\ket{\psi_i}$. If we apply our error detection procedure, we can filter out certain states, such as those with incorrect particle number parities. This results in a new density matrix, 
\begin{equation}
    \rho' = \sum_i p_i' \ket{\psi_i}\bra{\psi_i}.
\end{equation}
The energy of the system is given by 
\begin{equation}
    \mathrm{Tr}(H\rho') \geq E_g,
\end{equation}
as $\bra{\psi_i}H\ket{\psi_i} \geq E_g$ for all normalised wavefunctions $\ket{\psi_i}$.
Equality only holds when the filtering procedure is $100~\%$ effective, and the system is in the ground state. Consequently, the variational principle is not violated by our filtering procedure.\\

\section{Alternative stabiliser-VQE circuit}\label{stabiliserVQEcircuitproof}

Here we provide more detail on the alternative circuit mentioned in the main text, which can be used to reduce noise due to readout errors. We first show that the circuit given by Fig.~\ref{stabiliser}~\cite{ekert2002direct, kitaev1995phase} gives the desired outcome for the VQE; $\bra{\psi} h_j \ket{\psi}$. Stepping through the circuit, we find that:

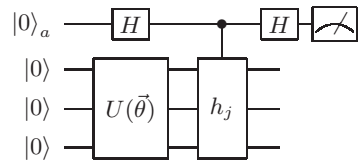
\begin{figure}[hbt]
	\begin{align*}
\Qcircuit @C=0.6em @R=.7em {
\lstick{\ket{0}_a}&\qw& \gate{H}&\qw&\ctrl{1}&\gate{H}&\meter\\
\lstick{\ket{0}}&\qw& \multigate{2}{U(\vec{\theta})}&\qw &\multigate{2}{h_j}&\qw\\
\lstick{\ket{0}}&\qw& \ghost{U(\vec{\theta})}&\qw& \ghost{h_j}&\qw\\
\lstick{\ket{0}}&\qw& \ghost{U(\vec{\theta})}&\qw& \ghost{h_j}&\qw\\
}
\end{align*}
\caption{The alternative method of performing a single measurement of term $h_j$ in the Hamiltonian. The ansatz circuit $U(\vec{\theta})$ creates a physical state; $\ket{\psi}=\ket{\psi(\vec{\theta})} = U(\vec{\theta})\ket{\bar{0}}$. The controlled-$h_j$ gate is easy to realise in practice, as $h_j$ is a product of Pauli terms on different qubits, so it can be implemented as a sequence of controlled Pauli gates.
}\label{stabiliser}
\end{figure}

\begin{gather}
		\ket{\bar{0}} \ket{0}_a \rightarrow \frac{1}{\sqrt{2}} ( \ket{\psi} \ket{0}_a + \ket{\psi} \ket{1}_a) \\ \notag
		\rightarrow \frac{1}{\sqrt{2}} ( \ket{\psi} \ket{0}_a + h_j \ket{\psi} \ket{1}_a) \\ \notag
		\rightarrow \frac{1}{2} [ \ket{0}_a (\ket{\psi} + h_j \ket{\psi})  +  \ket{1}_a (\ket{\psi} - h_j \ket{\psi})] = \ket{\phi}\\  \notag
\notag
\end{gather}

Measuring the ancilla in the computational basis gives
\begin{widetext}
\begin{gather}
		\bra{\phi} Z_a \ket{\phi} = \frac{1}{4}(2\langle \psi | \psi \rangle + 2 \bra{\psi} h_j \ket{\psi}) - \frac{1}{4}(2\langle \psi | \psi \rangle - 2 \bra{\psi} h_j \ket{\psi}) = \bra{\psi} h_j \ket{\psi},
\end{gather}
\end{widetext}
the same result as the conventional VQE circuit, as required. The circuit has obtained this result by performing measurement of a single qubit each time, which we will show reduces noise due to readout errors when compared with the normal direct measurement VQE protocol.

We can also use this circuit to extract parity information about the qubits `for free'. When the circuit and measurement are performed, the ancilla will either be measured in the $\ket{0}_a$ or the $\ket{1}_a$ state. The register is correspondingly in either 
\begin{gather}
	\ket{0}_a \rightarrow \ket{\phi_R} = \frac{\ket{\psi} + h_j \ket{\psi}}{\sqrt{2(1+\bra{\psi} h_j \ket{\psi})}} \\ 
	\ket{1}_a \rightarrow \ket{\phi_R} = \frac{\ket{\psi} - h_j \ket{\psi}}{\sqrt{2(1-\bra{\psi} h_j \ket{\psi})}}
\end{gather}
which we write as $\ket{\phi_R^\pm}$. \\

\begin{figure}[hbt]
	\begin{align*}
\Qcircuit @C=0.6em @R=.7em {
\lstick{\ket{0}_a}&\qw& \gate{H}&\qw&\ctrl{1}&\gate{H}&\meter\\
\lstick{\ket{0}}&\qw& \multigate{2}{U(\vec{\theta})}&\qw&\multigate{2}{h_j}&\qw&\qw&\qw&\multigate{2}{\langle \hat{P} \rangle}\\
\lstick{\ket{0}}&\qw& \ghost{U(\vec{\theta})}&\qw& \ghost{h_j}&\qw&\qw&\qw&\ghost{\langle \hat{P} \rangle}\\
\lstick{\ket{0}}&\qw& \ghost{U(\vec{\theta})}&\qw& \ghost{h_j}&\qw&\qw&\qw&\ghost{\langle \hat{P} \rangle}\\
}
\end{align*}
\caption{We perform a parity measurement on the register after the ancilla has been measured in order to detect errors.
}\label{StabVQE}
\end{figure}
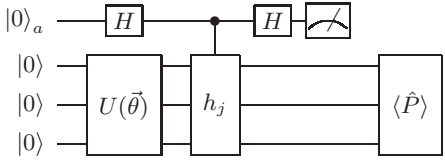

In the Jordan-Wigner encoding, the Pauli terms in the Hamiltonian, $h_j$, contain an even number of $X$ and $Y$ operators, and thus conserve electron number parity. Denoting the parity operator by $\hat{P}$, we find that $\hat{P} \ket{\psi} = (-1)^{N} \ket{\psi}$ and $\hat{P} h_j\ket{\psi} = (-1)^{N} h_j\ket{\psi}$, where $N$ is the number of electrons in the molecule. As a result, the register state is an eigenstate of the parity operator, and so we can measure this stabiliser quantity, as shown in Fig.~\ref{StabVQE}. If a single bit-flip error has occured, this will change the measured parity. If $\langle \hat{P} \rangle \neq (-1)^{N}$, we can discard the $h_j$ measurement result.

In the limit of small error rates, we consider that at most a single error happens. We assume here that no gate errors occur in the circuit, but a single readout error occurs. Using the circuit in Fig.~\ref{StabVQE}, the probability that the error happens on the ancilla qubit is $1/M$ where $M$ is the total number of qubits. If the error occurs on the ancilla qubit, its value will be corrupted. In contrast, the parity measurement will yield the correct result. This means we will not filter the corrupted result, reducing the accuracy of our expectation value measurement. In the other $(M-1)/M$ cases, the readout error occurs on one of the register qubits. This causes us to incorrectly assert that the state $\ket{\psi}$ is not a physical state, and therefore wrongly filter it out. However, our measured expectation value is not degraded much by this. When the number of qubits is large, the readout errors are much more likely to occur on the register qubits, and so we are still able to obtain an accurate expectation value.

In contrast, when considering the circuit shown in Fig.~1 in the main text, we find that the opposite is true. If only a single readout error occurs, it is much more likely to occur on the register qubits than on the ancilla. As a result, we are likely to accept a corrupted state, producing a less accurate expectation value. While this sounds like a significant disadvantage, we note that the reality is less clear cut. Readout error rates are currently at around 1~\%~\cite{preskill2018quantum} for superconducting qubits, higher than the lowest two qubit gate error rates. However, there are typically far more two qubit gates in the circuit than qubits to be measured, and so we expect gate errors to dominate. Moreover, this circuit variant is only capable of checking the total electron parity, and not the spin-up or down parities.\\

We can also apply this circuit to variational algorithms which simulate real~\cite{Li2017} and imaginary~\cite{mcardle2018variational} time evolution, as well as quantum gradient finding~\cite{romero2017strategies}. These algorithms use a circuit similar to the one shown in Fig.~\ref{stabiliser} to probe the derivative of an ansatz state with respect to one of the parameters. They leave the register in the state

\begin{gather}
	\ket{\phi_{\pm}} = \frac{1}{\sqrt{k_\pm}} (\ket{\psi(\vec{\theta})} \pm e^{i\theta} \sqrt{N} h_j \ket{\frac{\partial \psi(\vec{\theta})}{\partial \theta_i}})
\end{gather}
where $\ket{\psi(\vec{\theta})}$ is the ansatz state, $k_\pm$ is a normalisation constant, $h_j$ is a Hamiltonian term, $N$ is a normalisation constant for the derivative state vector, and $\theta=0, \pi$ depending on the whether the simulation is of real time, imaginary time, or gradient finding. 

If an ansatz is used which constructs physical trial states, then the state $\ket{\psi(\vec{\theta})}$ has fixed particle number and spin. As discussed above, the Hamiltonian operators $h_j$ are particle number parity conserving. Consequently, we only require that $\ket{\frac{\partial \psi(\vec{\theta})}{\partial \theta_i}}$ has a conserved particle number parity. This is the case for both UCC and Hamiltonian variational ansatze. Consequently, the qubit register is left in a state which has a fixed particle number parity, in the absence of errors. The particle number parity of the register qubits can then be checked, to detect possible errors. This error detection can be considered `free', as no additional ancilla qubits or circuit repititions are required, beyond what is needed for the quantum gradient finding circuit itself.

\section{Simulation details}
\subsection{UCC ansatz and the hydrogen molecule}\label{UCCandH2}

When solving the electronic structure problem for the hydrogen molecule in a minimal basis, we construct four molecular orbitals,
\begin{equation}
	\begin{aligned}
		|\sigma_{g \uparrow} \rangle = \frac{1}{\sqrt{S}}(| 1s_{A \uparrow} \rangle + | 1s_{B \uparrow} \rangle), \quad |\sigma_{g \downarrow} \rangle = \frac{1}{\sqrt{S}}(| 1s_{A \downarrow} \rangle + | 1s_{B \downarrow} \rangle), \\
		 |\sigma_{u \uparrow} \rangle = \frac{1}{\sqrt{S}}(| 1s_{A \uparrow} \rangle - | 1s_{B \uparrow} \rangle), \quad  |\sigma_{u \downarrow} \rangle = \frac{1}{\sqrt{S}}(| 1s_{A \downarrow} \rangle - | 1s_{B \downarrow} \rangle),
	\end{aligned}
\end{equation}
where $S$ is a normalisation factor, and $A$ and $B$ denote which of the two protons the $1s$ orbital is centred on. We can write a Jordan-Wigner (JW) mapped state vector as
\begin{equation}
	\begin{aligned}
		\ket{\psi} = 
		\ket{f_{\sigma_{u \downarrow}}, f_{\sigma_{g \downarrow}}, f_{\sigma_{u \uparrow}}, f_{\sigma_{g \uparrow}}},
	\end{aligned}
\end{equation}
where $f_i=1$ if spin orbital $i$ is occupied, and $f_i=0$ if spin orbital $i$ is unoccupied. Using the JW encoding, the 4 qubit Hamiltonian for H$_2$, given by
\begin{widetext}
\begin{equation}
	\begin{aligned}
	H =&~h_0 I + h_1 Z_0 + h_2 Z_1 + h_3 Z_2 + h_4 Z_3 + h_5 Z_0 Z_1 + h_6 Z_0 Z_2 + h_7 Z_1 Z_2 + h_8 Z_0 Z_3 + \\
		&~h_9 Z_1 Z_3 + h_{10} Z_2 Z_3 + h_{11} Y_0 Y_1 X_2 X_3 + h_{12} X_0 X_1 X_2 X_3 + h_{13} Y_0 Y_1 Y_2 Y_3 + h_{14} X_0 X_1 Y_2 Y_3.
	\end{aligned}
\end{equation}
\end{widetext}
We obtained the numerical values of $h_i$ using OpenFermion~\cite{openfermion}. The HF state for H$_2$ is given by
\begin{equation}
	\ket{\psi_{\mathrm{HF}}^{\mathrm{H_2}}} = \ket{0101}. 
\end{equation}
The most general state for H$_2$ (with the same spin and electron number as the HF state) is given by
\begin{equation}
	\ket{\psi^{\mathrm{H_2}}} = \alpha \ket{0101} + \beta \ket{1010} + \gamma \ket{1001} + \delta \ket{0110}.
\end{equation}

\begin{figure}[hbt]
\centering
\begin{align*}
\Qcircuit @C=0.8em @R=.7em {
\lstick{ \sigma_{g\uparrow} : \ket{1}}&\gate{H}& \ctrl{1} & \qw & \qw &\qw & \qw & \qw & \ctrl{1} & \gate{H} & \qw	\\
\lstick{\sigma_{u\uparrow} :  \ket{0}}&\gate{R_x(\frac{\pi}{2})}&	\targ &	\ctrl{1} & \qw &\qw & \qw & \ctrl{1} & \targ	&\gate{R_x(-\frac{\pi}{2})} &\qw\\
\lstick{\sigma_{g\downarrow} :  \ket{1}}&\gate{H}& \qw & \targ &	\ctrl{1} 	&\qw &	\ctrl{1} & \targ & \qw & \gate{H} &\qw	\\
\lstick{\sigma_{u\downarrow} :  \ket{0}}&\gate{H}&	\qw &\qw & \targ & \gate{R_z(-\theta)} & \targ &	\qw &\qw & \gate{H} & \qw	\\
} \\ \\
\Qcircuit @C=0.8em @R=.7em {
\lstick{}&\gate{H}& \ctrl{1} & \qw & \qw &\qw & \qw & \qw & \ctrl{1} & \gate{H} & \qw	\\
\lstick{}&\gate{H}&	\targ &	\ctrl{1} & \qw &\qw & \qw & \ctrl{1} & \targ	&\gate{H} &\qw\\
\lstick{}&\gate{H}& \qw & \targ &	\ctrl{1} 	&\qw &	\ctrl{1} & \targ & \qw & \gate{H} &\qw	\\
\lstick{}&\gate{R_x(\frac{\pi}{2})}&	\qw &\qw & \targ & \gate{R_z(-\theta)} & \targ &	\qw &\qw & \gate{R_x(-\frac{\pi}{2})} & \qw	\\
} \\ \\
\Qcircuit @C=0.8em @R=.7em {
\lstick{}&\gate{H}& \ctrl{1} & \qw & \qw &\qw & \qw & \qw & \ctrl{1} & \gate{H} & \qw	\\
\lstick{}&\gate{R_x(\frac{\pi}{2})}&	\targ &	\ctrl{1} & \qw &\qw & \qw & \ctrl{1} & \targ	&\gate{R_x(-\frac{\pi}{2})} &\qw\\
\lstick{}&\gate{R_x(\frac{\pi}{2})}& \qw & \targ &	\ctrl{1} 	&\qw &	\ctrl{1} & \targ & \qw & \gate{R_x(-\frac{\pi}{2})} &\qw	\\
\lstick{}&\gate{R_x(\frac{\pi}{2})}&	\qw &\qw & \targ & \gate{R_z(-\theta)} & \targ &	\qw &\qw & \gate{R_x(-\frac{\pi}{2})} & \qw	\\
} \\ \\
\Qcircuit @C=0.8em @R=.7em {
\lstick{}&\gate{R_x(\frac{\pi}{2})}& \ctrl{1} & \qw & \qw &\qw & \qw & \qw & \ctrl{1} & \gate{R_x(-\frac{\pi}{2})} & \qw	\\
\lstick{}&\gate{R_x(\frac{\pi}{2})}&	\targ &	\ctrl{1} & \qw &\qw & \qw & \ctrl{1} & \targ	&\gate{R_x(-\frac{\pi}{2})} &\qw\\
\lstick{}&\gate{H}& \qw & \targ &	\ctrl{1} 	&\qw &	\ctrl{1} & \targ & \qw & \gate{H} &\qw	\\
\lstick{}&\gate{R_x(\frac{\pi}{2})}&	\qw &\qw & \targ & \gate{R_z(-\theta)} & \targ &	\qw &\qw & \gate{R_x(-\frac{\pi}{2})} & \qw	\\
} \\ \\
\Qcircuit @C=0.8em @R=.7em {
\lstick{}&\gate{R_x(\frac{\pi}{2})}& \ctrl{1} & \qw & \qw &\qw & \qw & \qw & \ctrl{1} & \gate{R_x(-\frac{\pi}{2})} & \qw	\\
\lstick{}&\gate{R_x(\frac{\pi}{2})}&	\targ &	\ctrl{1} & \qw &\qw & \qw & \ctrl{1} & \targ	&\gate{R_x(-\frac{\pi}{2})} &\qw\\
\lstick{}&\gate{R_x(\frac{\pi}{2})}& \qw & \targ &	\ctrl{1} 	&\qw &	\ctrl{1} & \targ & \qw & \gate{R_x(-\frac{\pi}{2})} &\qw	\\
\lstick{}&\gate{H}&	\qw &\qw & \targ & \gate{R_z(\theta)} & \targ &	\qw &\qw & \gate{H} & \qw	\\
} \\ \\
\Qcircuit @C=0.8em @R=.7em {
\lstick{}&\gate{R_x(\frac{\pi}{2})}& \ctrl{1} & \qw & \qw &\qw & \qw & \qw & \ctrl{1} & \gate{R_x(-\frac{\pi}{2})} & \qw	\\
\lstick{}&\gate{H}&	\targ &	\ctrl{1} & \qw &\qw & \qw & \ctrl{1} & \targ	&\gate{H} &\qw\\
\lstick{}&\gate{R_x(\frac{\pi}{2})}& \qw & \targ &	\ctrl{1} 	&\qw &	\ctrl{1} & \targ & \qw & \gate{R_x(-\frac{\pi}{2})} &\qw	\\
\lstick{}&\gate{R_x(\frac{\pi}{2})}&	\qw &\qw & \targ & \gate{R_z(\theta)} & \targ &	\qw &\qw & \gate{R_x(-\frac{\pi}{2})} & \qw	\\
} \\ \\
\end{align*}
\caption{The first half of the complete circuit used in our simulations to implement the UCCSD operator for H$_2$. The $R_x(\frac{\pi}{2})$ and $H$ gates rotate the basis such that the exponentiated operator applied to the corresponding qubit is either $Y$ or $X$, respectively.} \label{UCCH2a}
\end{figure}
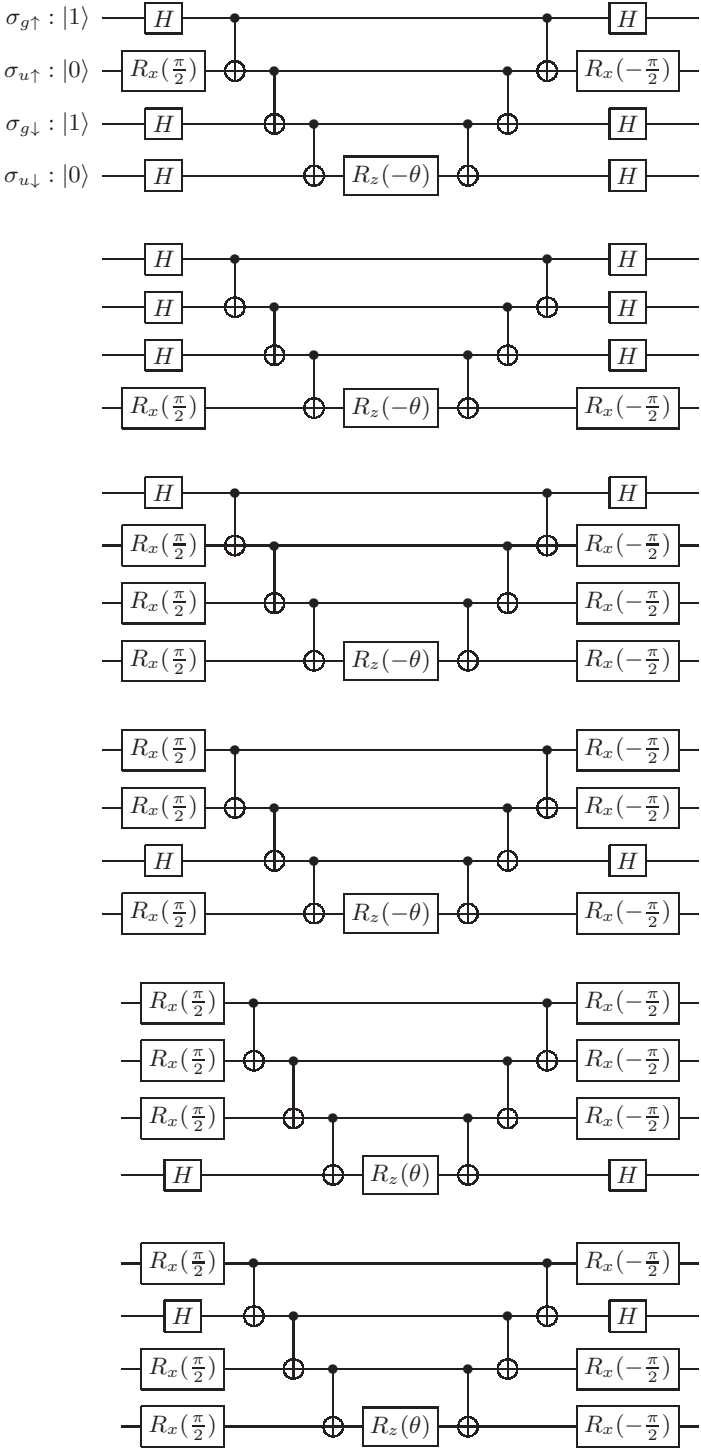

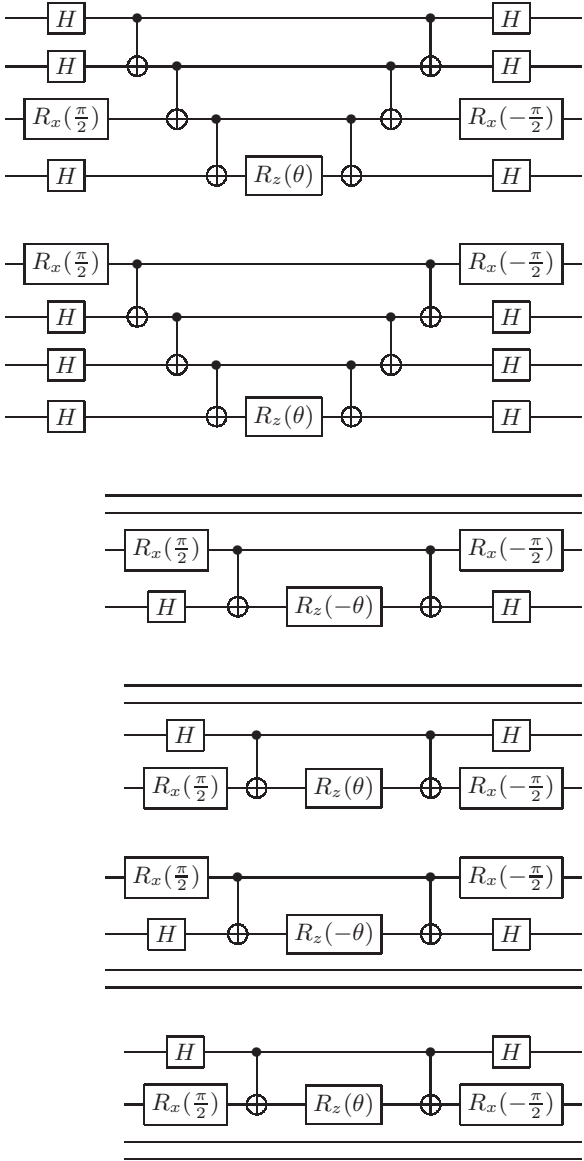
\begin{figure}[hbt]
\centering
\begin{align*}
\Qcircuit @C=0.8em @R=.7em {
\lstick{}&\gate{H}& \ctrl{1} & \qw & \qw &\qw & \qw & \qw & \ctrl{1} & \gate{H} & \qw	\\
\lstick{}&\gate{H}&	\targ &	\ctrl{1} & \qw &\qw & \qw & \ctrl{1} & \targ	&\gate{H} &\qw\\
\lstick{}&\gate{R_x(\frac{\pi}{2})}& \qw & \targ &	\ctrl{1} 	&\qw &	\ctrl{1} & \targ & \qw & \gate{R_x(-\frac{\pi}{2})} &\qw	\\
\lstick{}&\gate{H}&	\qw &\qw & \targ & \gate{R_z(\theta)} & \targ &	\qw &\qw & \gate{H} & \qw	\\
} \\ \\
\Qcircuit @C=0.8em @R=.7em {
\lstick{}&\gate{R_x(\frac{\pi}{2})}& \ctrl{1} & \qw & \qw &\qw & \qw & \qw & \ctrl{1} & \gate{R_x(-\frac{\pi}{2})} & \qw	\\
\lstick{}&\gate{H}&	\targ &	\ctrl{1} & \qw &\qw & \qw & \ctrl{1} & \targ	&\gate{H} &\qw\\
\lstick{}&\gate{H}& \qw & \targ &	\ctrl{1} 	&\qw &	\ctrl{1} & \targ & \qw & \gate{H} &\qw	\\
\lstick{}&\gate{H}&	\qw &\qw & \targ & \gate{R_z(\theta)} & \targ &	\qw &\qw & \gate{H} & \qw	\\
} \\ \\
\Qcircuit @C=0.8em @R=.7em {
\lstick{}&\qw & \qw & \qw &\qw & \qw & \qw  & \qw & \qw	\\
\lstick{}&\qw & \qw & \qw &\qw & \qw & \qw  & \qw & \qw	\\
\lstick{}&\gate{R_x(\frac{\pi}{2})}&	\ctrl{1} & \qw &\qw & \qw & \ctrl{1} 	&\gate{R_x(-\frac{\pi}{2})} &\qw\\
\lstick{}&\gate{H} & \targ & \qw & \gate{R_z(-\theta)}  & \qw & \targ & \gate{H} & \qw 	\\
} \\ \\
\Qcircuit @C=0.8em @R=.7em {
\lstick{}&\qw & \qw & \qw &\qw & \qw & \qw  & \qw & \qw	\\
\lstick{}&\qw & \qw & \qw &\qw & \qw & \qw  & \qw & \qw	\\
\lstick{}&\gate{H}&	\ctrl{1} & \qw &\qw & \qw & \ctrl{1} 	&\gate{H} &\qw\\
\lstick{}&\gate{R_x(\frac{\pi}{2})} & \targ & \qw & \gate{R_z(\theta)}  & \qw & \targ & \gate{R_x(-\frac{\pi}{2})} & \qw 	\\
} \\ \\
\Qcircuit @C=0.8em @R=.7em {
\lstick{}&\gate{R_x(\frac{\pi}{2})}&	\ctrl{1} & \qw &\qw & \qw & \ctrl{1} 	&\gate{R_x(-\frac{\pi}{2})} &\qw\\
\lstick{}&\gate{H} & \targ & \qw & \gate{R_z(-\theta)}  & \qw & \targ & \gate{H} & \qw 	\\
\lstick{}&\qw & \qw & \qw &\qw & \qw & \qw  & \qw & \qw	\\
\lstick{}&\qw & \qw & \qw &\qw & \qw & \qw  & \qw & \qw	\\
} \\ \\
\Qcircuit @C=0.8em @R=.7em {
\lstick{}&\gate{H}&	\ctrl{1} & \qw &\qw & \qw & \ctrl{1} 	&\gate{H} &\qw\\
\lstick{}&\gate{R_x(\frac{\pi}{2})} & \targ & \qw & \gate{R_z(\theta)}  & \qw & \targ & \gate{R_x(-\frac{\pi}{2})} & \qw 	\\
\lstick{}&\qw & \qw & \qw &\qw & \qw & \qw  & \qw & \qw	\\
\lstick{}&\qw & \qw & \qw &\qw & \qw & \qw  & \qw & \qw	\\
}
\end{align*}
\caption{The second half of the complete circuit used in our simulations to implement the UCCSD operator for H$_2$. The $R_x(\frac{\pi}{2})$ and $H$ gates rotate the basis such that the exponentiated operator applied to the corresponding qubit is either $Y$ or $X$, respectively.} \label{UCCH2b}
\end{figure}

We can construct such a state using the singlet unitary coupled cluster (UCC) ansatz, considering single and double excitations above the HF state (UCCSD). The UCCSD operator is given by
\begin{equation}
	\begin{aligned}
		U &= e^{(T_1 - T_1^\dag) + (T_2 - T_2^\dag)}, \\ 
	\end{aligned}
\end{equation}
where
\begin{equation}
	\begin{aligned}
		T_1 &=  \sum_{i \in virt, \alpha \in occ} t_{i \alpha} a^\dag_i a_\alpha, \\ 
		T_2 &= \sum_{i, j \in virt, \alpha, \beta \in occ} t_{i j \alpha \beta}  a^\dag_i a^\dag_j a_\alpha a_\beta, \\ 
	\end{aligned}
\end{equation}
and $occ$ are occupied orbitals, $virt$ are initially unoccupied orbitals in the reference state, and $t_{i \alpha}$ and $t_{ij\alpha \beta}$ are variational parameters to be optimised. For H$_2$, the singlet UCCSD operator takes the form
\begin{equation}\label{ucceq}
	\begin{aligned}
		U = e^{t_{10}(a_1^\dag a_0 - a_0^\dag a_1) + t_{32}(a_3^\dag a_2 - a_2^\dag a_3) + t_{3120}(a_3^\dag a_1^\dag a_2 a_0 - a_0^\dag a_2^\dag a_1 a_3)}. 
	\end{aligned}
\end{equation}
Splitting this operator up using Trotterization (with a single Trotter step) and using the JW encoding, we find that
\begin{widetext}
\begin{gather}\label{ucceq2}
		(a_1^\dag a_0 - a_0^\dag a_1) = \frac{i}{2}(X_1 Y_0 - Y_1 X_0) \notag \\
		(a_3^\dag a_2 - a_2^\dag a_3) = \frac{i}{2}(X_3 Y_2 - Y_3 X_2) \\
		(a_3^\dag a_1^\dag a_2 a_0 - a_0^\dag a_2^\dag a_1 a_3) =\notag \\ 
		\frac{i}{8}(X_3 X_2 Y_1 X_0 + Y_3 X_2 X_1 X_0 + Y_3 Y_2 Y_1 X_0 + Y_3 X_2 Y_1 Y_0 - X_3 Y_2 Y_1 Y_0 - Y_3 Y_2 X_1 Y_0 - X_3 Y_2 X_1 X_0 - X_3 X_2 X_1 Y_0). \notag
\end{gather}
\end{widetext}
We can apply this operator using the circuit shown in Fig.~\ref{UCCH2a} and Fig.~\ref{UCCH2b}, generated using the circuit constructions in Ref.~\cite{whitfield2011simulation}.

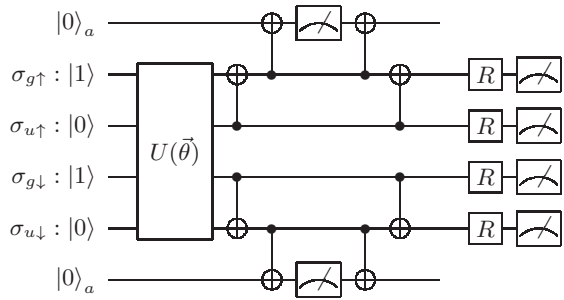
\begin{figure}[hbt]
	\begin{align*}
\Qcircuit @C=0.6em @R=.7em {
\lstick{\ket{0}_a} &\qw &\qw &\qw &\targ &\meter &\targ &\qw &\qw &\qw \\
\lstick{\sigma_{g\uparrow} : \ket{1}}&\qw& \multigate{3}{U(\vec{\theta})}&\targ &\ctrl{-1}&\qw&\ctrl{-1}&\targ&\qw&\qw&\qw&\gate{R}&\meter \\
\lstick{\sigma_{u\uparrow} : \ket{0}}&\qw& \ghost{U(\vec{\theta})}&\ctrl{-1} &\qw&\qw&\qw&\ctrl{-1}&\qw&\qw&\qw&\gate{R}&\meter \\
\lstick{\sigma_{g\downarrow} : \ket{1}}&\qw& \ghost{U(\vec{\theta})}&\ctrl{1}& \qw&\qw&\qw&\ctrl{1}&\qw&\qw&\qw&\gate{R}&\meter \\
\lstick{\sigma_{u\downarrow} : \ket{0}}&\qw& \ghost{U(\vec{\theta})}&\targ& \ctrl{1}&\qw&\ctrl{1}&\targ&\qw&\qw&\qw&\gate{R}&\meter \\
\lstick{\ket{0}_a}&\qw& \qw&\qw&\targ&\meter&\targ & \qw&\qw&\qw \\
}
	\end{align*}
\caption{A circuit using a local gateset to measure both the spin-up parity and spin-down parity. The $R$ gates implement the single qubit basis rotations required to measure the desired Hamiltonian string.}\label{paritychecklocal}
\end{figure}

The parity check circuit used in our error-mitigated simulation of the Hydrogen molecule is shown in Fig.~\ref{paritychecklocal}. We restricted ourselves to a linear nearest-neighbour connectivity, in order to lower bound the efficacy of our method. The electron spin number parity checks could be implemented by using a series of SWAP gates to move the ancilla qubits along the register, such that they are next to every register qubit once. This can be achieved in depth $O(M)$. Instead, it is more efficient to use the procedure shown in Fig.~\ref{paritychecklocal}, which effectively `passes' the parity information along the register to the ancilla, before uncomputing the procedure to ensure reversibility. This circuit has already been suggested for efficient stabiliser checking in Ref.~\cite{bonet2018mitigation}.

\subsection{Number of measurements and error analysis}
Our numerical simulations were designed to mimic the actions of an experimentalist; the expectation value of each term in the Hamiltonian was found by repeating the circuit and measurement procedure many times.

The number of measurements required was set by the desired precision. Measurements were distributed optimally among the different Hamiltonian terms~\cite{VQETheoryNJP, romero2017strategies}, such that the number of measurements for each term was proportional to its strength. The standard error in each measurement is then given as follows.

The standard error in the mean is given by~\cite{hughesandhase} 
\begin{equation}
    \alpha = \frac{\sigma}{\sqrt{N}}
\end{equation}
where $N$ is the number of measurements performed, and $\sigma$ is the standard deviation of the result. The standard deviation of a measurement of one of the Pauli strings in the Hamiltonian, $\hat{h}_j$, is given by
\begin{equation}
    \sigma_{h_j} = \sqrt{\bra{\psi} \hat{h}_j^2 \ket{\psi} - \bra{\psi} \hat{h}_j \ket{\psi}^2} = \sqrt{1 - \bra{\psi} \hat{h}_j \ket{\psi}^2} \leq 1.
\end{equation}
The standard error in the energy measurement is then upper bounded by
\begin{equation}\label{errorinH}
    \alpha_E =\sqrt{\sum_i \alpha_{h_i}^2} = \sqrt{\sum_i |g_i|^2 \frac{\sigma_{h_j}^2}{N_i}} \leq \sqrt{\sum_i \frac{|g_i|^2}{N_i}},
\end{equation}
where $N_i$ is the number of measurements used for each term in the Hamiltonian, and $g_i$ is the coefficient of term $i$ in the Hamiltonian.
We distribute our measurements optimally~\cite{VQETheoryNJP,romero2017strategies}, setting
\begin{equation}\label{numberofmeasurements}
    N_i = \frac{|g_i|}{g_{\mathrm{max}}}k
\end{equation}
where $g_{\mathrm{max}}$ is the largest coefficient in the Hamiltonian, and $k$ is the number of measurements allocated to the largest term in the Hamiltonian. We substitute this expression into Eq.~\ref{errorinH},
\begin{equation}\label{energyerror}
    \alpha_E \leq \sqrt{\sum_i \frac{|g_i|g_{\mathrm{max}}}{k}}
\end{equation}
Solving for $k$, and substituting back into Eq.~\ref{numberofmeasurements} provides an expression for the number of measurements required per term, as a function of the standard error
\begin{equation}
    N_i = \frac{|g_i| \sum_j |g_j|}{\alpha_E^2}. 
\end{equation}
The total number of measurements is then 
\begin{equation}
    N = \frac{(\sum_i |g_i|)^2}{\alpha_E^2}. 
\end{equation}\\

When extrapolation is performed, the standard error is increased by a factor that depends on the `stretch-factor', $\lambda$, used in the extrapolation. When performing extrapolation, we used the same total number of samples for each expectation value, which were divided equally between two points for a linear extrapolation. The extrapolated value is given by
\begin{equation}
    O_{\mathrm{extrap}} = \frac{\lambda O(\epsilon) - O(\lambda \epsilon)}{\lambda - 1}. 
\end{equation}
As such, the standard error was increased by a factor of
\begin{equation}
    \omega = \frac{\sqrt{2(\lambda^2 + 1)}}{(\lambda - 1)}.
\end{equation}
The stretch factor used in the simulations shown in Fig.4 depended on the error rate, $\epsilon$, as 
\begin{equation}
    \lambda = 1 + \frac{0.001}{\epsilon}.
\end{equation}
The stretch factor used in the simulations shown in Fig.5 was 
\begin{equation}
    \lambda = 1.5~.
\end{equation}

For the results presented in Fig.~5 in the main text, we calculated the standard error using the expectation values $\sqrt{1 - \bra{\psi} \hat{h}_j \ket{\psi}^2}$, rather than the loose upper bound. The resulting standard errors in the results were: Detection + Extrapolation = [0.05, 0.07, 0.04, 0.04, 0.04] mhartree, Extrapolation = [0.05, 0.07, 0.04, 0.05, 0.04] mhartree.

For the results presented in Fig.~4 in the main text, we used the loose upper bound for the standard error, obtained as described above.

\bibliographystyle{apsrev4-1} 
\bibliography{ChemistryReviewBib}

\end{document}